\documentclass[aps,prb,twocolumn,superscriptaddress,email]{revtex4}  
\usepackage[pdftex]{graphicx}  
\usepackage{dcolumn}   
\usepackage{bm}        
\usepackage{amssymb}   

\usepackage{amsmath}
\usepackage[latin1]{inputenc}

\usepackage{color}

\begin{document}
\title{Dissipative dynamics of a solid-state qubit coupled to surface plasmons: from
non-Markov to Markov regimes}
\author{A.Gonzalez-Tudela}
\email[Corresponding author: ]{alejandro.gonzalez@uam.es}
\affiliation{F\'{\i}sica Te\'orica de la Materia Condensada, Universidad Aut\'onoma de Madrid, 28049, Spain.}
\author{ F.J.Rodr\'{\i}guez}
\affiliation{Departamento de F\'{\i}sica, Universidad de los Andes.
A.A. 4976, Bogot\'a, Colombia}
\author{L.Quiroga}
\affiliation{Departamento de F\'{\i}sica, Universidad de los Andes.
A.A. 4976, Bogot\'a, Colombia}
\author{C. Tejedor}
\affiliation{F\'{\i}sica Te\'orica de la Materia Condensada, Universidad Aut\'onoma de Madrid, 28049, Spain.}
\date{\today}

\begin{abstract}
We theoretically study the dissipative dynamics of a quantum emitter
placed near the planar surface of a metal supporting surface plasmon
excitations. The emitter-metal coupling regime can be tuned by
varying some control parameters such as the qubit-surface separation
and/or the detuning between characteristic frequencies. By using a
Green's function approach jointly with a time-convolutionless master
equation, we analyze the non-Markovian dissipative features on the
qubit time evolution in two cases of interest: {\it i)} an undriven
qubit initially prepared in its excited state and {\it ii)} the
evolution towards a steady-state for a system driven by a laser
field. For weak to moderate qubit-metal coupling strength, and on
timescales large compared to the surface plasmon oscillation time, a
Markovian approximation for the master equation results to be
adequate to describe the qubit main optical properties: surface
enhancements of rate emission, optical spectra and time-dependent
photon-photon correlation functions. The qubit decay shows a crossover
passing from being purely dissipative for
small qubit-surface distances to plasmon emission for larger separations.
\end{abstract}

\maketitle

\section{Introduction\label{Intro}}

Surface plasmons (SP) on metals, a topic extensively studied from
many years ago\cite{raether88}, has recently received a strongly
renewed attention due to significant advances in new experimental
capabilities and numerical
developments\cite{barnes03,fj10}. Great attention has been focused
on the emerging field of quantum plasmonic with the goal of making
devices for quantum information processing\cite{chang06a,chang07a}
as single-photon transistor \cite{chang07b} or
lasers\cite{oulton09a}. As a requisite for this goal, a lot of
effort has been devoted to get coherent coupling between plasmons
and a quantum emitter made of a solid state qubit (SSQ) as, for instance,
a quantum dot, a single NV center or a single molecule among others.

Strong coupling signatures of SSQ and SP have
been experimentally observed both in spectroscopic as well as in
time-resolved studies. In spectroscopy, the
anticrossing between exciton and plasmon features in optical
spectra have already been reported in systems such as organic
semiconductors\cite{bellesa04a}, organic molecules placed in
subwavelength hole arrays\cite{dintinger05a}, metallic
nanowires\cite{akimov07a}, hybrid metal-semiconductor
nanostructures\cite{vasa08a} and even in carbon
nanotubes\cite{bondarev09a}. These anticrossings have been claimed
to be a manifestation of strong coupling between SSQ and SP. On the
other hand, ultrafast time-resolved signatures of strong coupling in SSQ-SP
systems have also been reported. An enhancement of several orders of magnitude
for the spontaneous emission rate in a time-resolved
photoluminescence measurement on a InGaN heterostructure close to a
silver thin layer has been reported \cite{neogi02a,okamoto05a}.
Additionally, recent experiments which operate simultaneously with
both Raman and fluorescence signals coming from a single molecule in
very close proximity to a metal surface have allowed the indirect
measurement of ultrafast ($\sim 25$ fs) dynamical features in such
SSQ-SP system\cite{galloway09a}.

On the theoretical side, some progress has been made to understand
SSQ-SP coupling in different geometries using different approaches.
The first attempts were devoted to computing the spontaneous
emission rate enhancement of an atom near an absorbing surface as
given by the atom self-energy in a near field limit\cite{yeung96}.
More recently, a hydrodynamic model\cite{trugler08a} has been used
to study a single molecule coupled to metallic nanoparticles. A
transfer matrix method has also been used for simulating
attenuated-reflection experiments\cite{tame08a}. However, the
experimental setup which has risen the highest interest has been the
quantum emitter coupled to a metallic
nanowire\cite{akimov07a,kolesov09a}, where the generation of a
single optical plasmon can be achieved. Several theoretical studies
on this system have considered the full quantum behavior of plasmon
modes \cite{klimov04,chen09a,dzsotjan10a}. In particular, some attention
has been devoted\cite{chen09a} to non-Markovian effects that can be
important in the SSQ-nanowire system because the
spectral density $J(\omega)$ (carefully discussed in the present
work) is highly structured due to a divergence at the edge of the
SP density of states.

\begin{figure}[ht]
  \centering
\includegraphics[width=.65\linewidth]{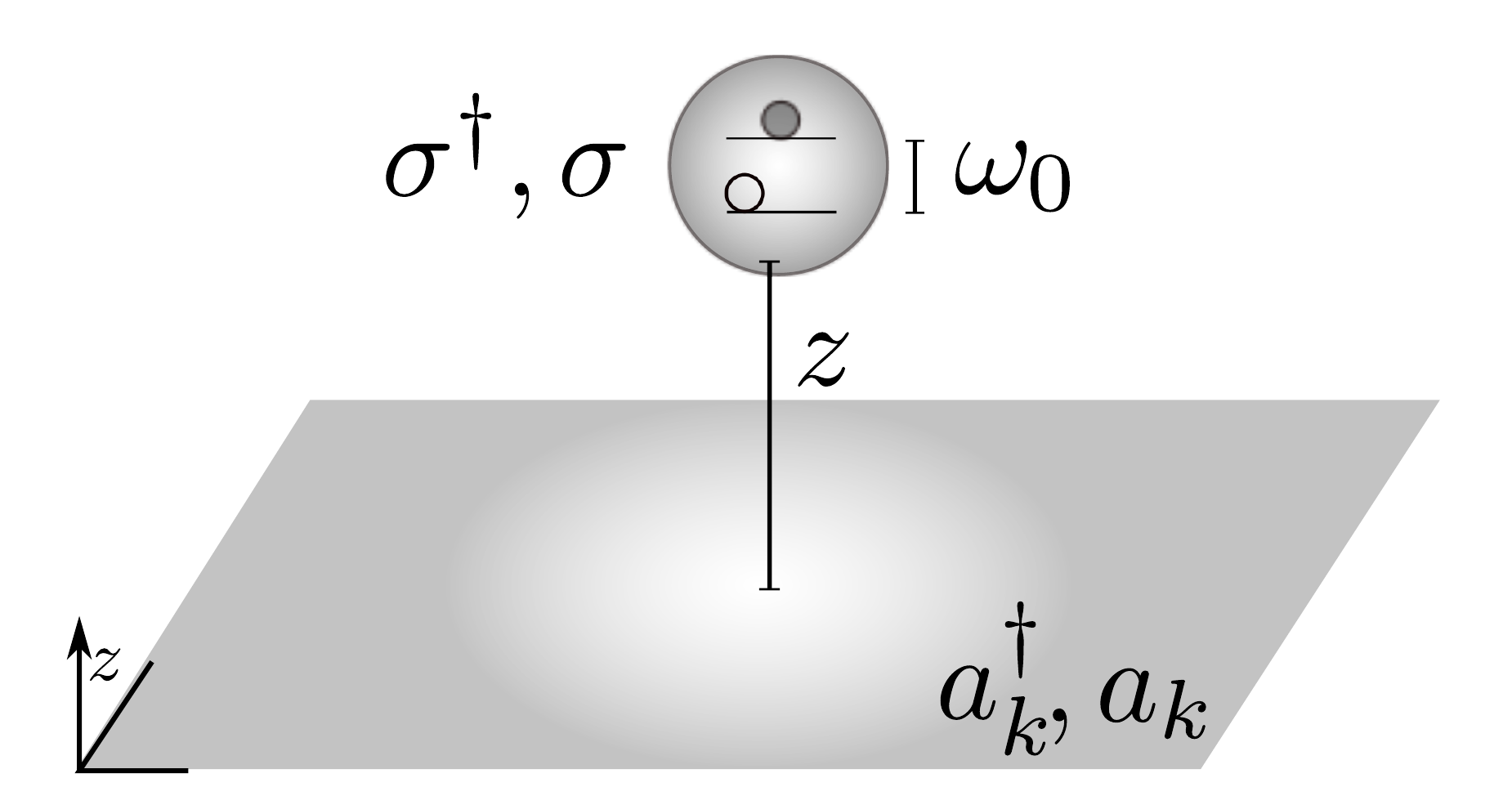}
  \caption{Schematic view of the system: a SSQ, with characteristic
  frequency $\omega_0$, is placed at a distance $z$ of an infinite
  planar metallic surface.}
  \label{fig:sistema}
\end{figure}

An open quantum system strongly coupled to a
reservoir displays a complex dynamics which, in general, requires a description
beyond simple Markovian theories\cite{breuer99c,ferney08a}. In order
to clarify the relevance of non-Markovian effects in SSQ-SP systems,
we concentrate in a quantum emitter close to a planar surface of a
dissipative metal, a system conceptually simpler than wires because
it only has a single band of plasmons with a density of states
having a singularity at a frequency $\omega_{sp}$.
We study the properties of the light emitted by the system
depicted in figure \ref{fig:sistema}: a SSQ close to the planar
metallic surface which supports a plasmon field as well as some dissipation
mechanism. Strong SSQ-SP coupling could
be expected when the qubit-surface distance is small compared with a
typical length scale as, for instance, the wavelength of the emitted light. In
order to understand the fundamental mechanisms of SSQ-SP strong
coupling, we restrict ourselves to consider just a single quantum
emitter. However, collective effects of many emitters coupled to
the same plasmon field have been recently proposed\cite{savasta10a}
as responsible for the detection of the Rabi vacuum splitting in
these systems.

We start by analyzing non-Markovian features in the SSQ-SP optical
features by using a time-convolutionless
approach\cite{breuerbook02}. We show that SSQ-SP dissipative
evolution is determined by the rapidly varying structure of the
reservoir spectral function at a frequency close to $\omega_{sp}$ of
the order of a few $eV$ in a normal metal. Therefore, the timescale
for non-Markovian effects reduces, at most, to a few hundreds times
$eV^{-1}$, i.e. typical times under a picosecond. Consequently, we
conclude that the observation of non-Markovian signatures in SSQ-SP
systems made with normal metals will indeed demand experimental set
ups at the edge front of present state-of-the-art ultrafast
technology. Beyond this short timescale, rate emissions stay
constant and a Markovian approach becomes adequate for computing
population dynamics, optical spectrum and second order coherence
function. Here, both a Markovian and a non-Markovian
analysis are applied to a SSQ-SP system under two different
excitation schemes: firstly, we analyze the spontaneous emission of
a SSQ initially prepared in the excited state. Secondly, the time
evolution of a SSQ initially in its ground state and driven by means
of a coherent laser field up to a stationary state is studied.

We take $\hbar=1$ along this paper which is organized as follows: in
Section \ref{sec:green} we introduce the Green's tensor of the layered
system and study its main properties, in particular the spectral
density function. In Section \ref{sec:tcl} the time-convolutionless
method is briefly reviewed and the non-Markovian effects on the
SSQ-SP system dynamics are considered. In Section \ref{sec:Markov}
we use the Markovian limit to calculate the optical properties of
the system. Finally, in Section \ref{sec:Summary} we summarize
our results and draw some conclusions.

\section{Green's tensor and spectral density\label{sec:green}}

Electrodynamics of a dissipative medium is described by the Green's
tensor  $\widehat{\mathbf{G}}(\mathbf{r},\mathbf{r}^{\prime},\omega)$ which
satisfies the Maxwell equation:
\begin{eqnarray}
 \label{Maxwell}
\left[ \left( \nabla \times \nabla \times \right)
-\frac{\omega^2}{c^2}\epsilon(\mathbf{r},\omega)\right] \widehat{\mathbf{G}}(\mathbf{r},\mathbf{r}^{\prime},\omega)=
\widehat{\mathbf{I}}\delta(\mathbf{r}-\mathbf{r}^{\prime}).
\end{eqnarray}
We study the system depicted in figure \ref{fig:sistema}: a SSQ in
the upper-half space is embedded within a dielectric matrix with a
dielectric function that can be taken as real and constant,
$\epsilon_1$, in the range of frequencies of interest. In the lower
half-space, $z<0$, a dissipative metal is characterized by a complex
dielectric function $\epsilon_2(\omega)$ that we take in a
renormalized Drude approximation:
\begin{equation}
 \label{dielectMetal}
\epsilon_2(\omega)=\epsilon_\infty \left(1-\frac{\omega_p^2}{\omega(\omega+i\gamma_p)}\right).
\end{equation}
$\epsilon_\infty$ is the high-frequency limit of the metal
dielectric function, $\omega_p$ is the bulk plasmon frequency and
$\gamma_p$ is the Landau damping constant.

\begin{figure}[ht]
\centering
\includegraphics[width=0.99\linewidth]{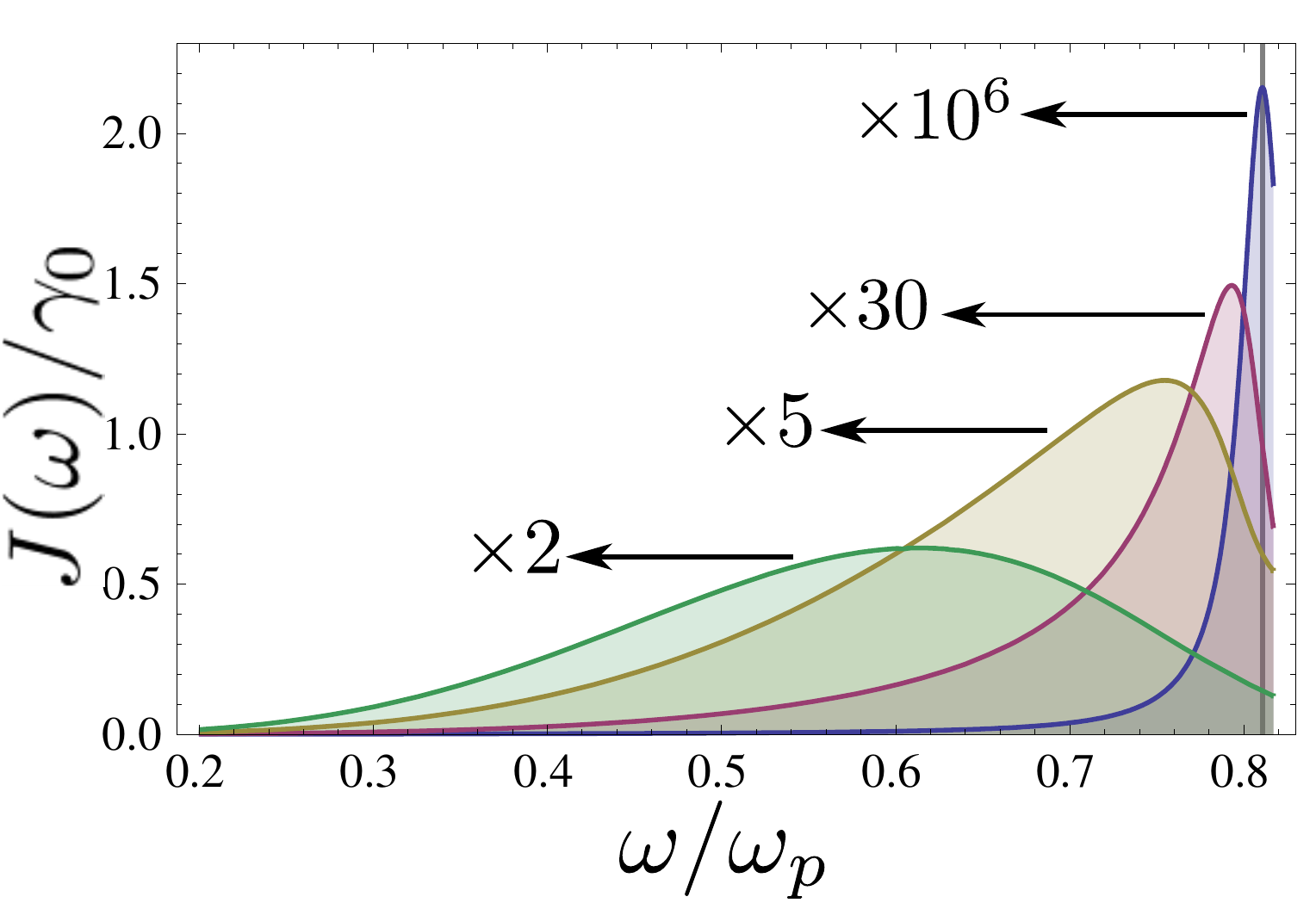}
  \caption{Spectral density $J(\omega)/\gamma_0$ for a detuning $\Delta=\bar{\omega}_{sp}-\bar{\omega}_0=0.1$ plotted for different qubit-surface separations: $\bar{z}=\omega_p
z/c=0.01$(blue), $0.32$(purple), $0.64$(yellow) and $1.42$(green).
Notice that as far as the SSQ approaches the surface, the spectral
density increases in roughly 6 orders of magnitude.}
  \label{fig:density}
\end{figure}

The Green's tensor for this layered geometry has two contributions:
\begin{eqnarray} \nonumber
\widehat{\mathbf{G}}(\mathbf{r},\mathbf{r}^{\prime},\omega)=\widehat{\mathbf{G}}_0(\mathbf{r},\mathbf{r}^{\prime},\omega)+
\widehat{\mathbf{G}}_R(\mathbf{r},\mathbf{r}^{\prime},\omega) \label{Eq:s17}
\end{eqnarray}
where the first term is the free-space solution given
by\cite{scheel08a}
\begin{eqnarray}
& & \widehat{\mathbf{G}}_0(\mathbf{r},\mathbf{r}^{\prime},\omega)= -\frac{1}{3q^2}\delta(\mathbf{R})\widehat{\mathbf{I}}
\nonumber  \\ & & +\frac{q}{4\pi}\left
[ \phi\left(\frac{1}{qR}\right)\widehat{\mathbf{I}}+\chi\left(\frac{1}{qR}\right)\frac{\mathbf{R}
\otimes \mathbf{R}}{R^2} \right ]e^{iqR} \label{Eq:s14}
\end{eqnarray}
with $q=\omega/c$, $\phi(x)=x+ix^2-x^3$, $\chi(x)=x+3ix^2-3x^3$,
$\widehat{\mathbf{I}}$ is the $3 \times 3$ identity matrix and
$\mathbf{R} \otimes \mathbf{R}$ represents the dyadic product
between the vectors $\mathbf{R}=\mathbf{r}-\mathbf{r}^{\prime}$. The
second term in Eq.(\ref{Eq:s17}),
$\widehat{\mathbf{G}}_R(\mathbf{r},\mathbf{r}^{\prime},\omega)$, is
the reflection Green's tensor with on-site nonzero components\cite{scheel08a}
\begin{widetext}\begin{eqnarray}
G_{R,zz}(\mathbf{r},\mathbf{r},\omega)&=&-\frac{c^2}{4\pi
\omega^2}\int_0^{\infty}dq \frac{q^3}{\kappa_1(q,\omega)} \left [
\frac{\epsilon_1\kappa_2(q,\omega)-\epsilon_2(\omega)\kappa_1(q,\omega)}{\epsilon_1\kappa_2(q,\omega)+\epsilon_2(\omega)\kappa_1(q,\omega)}\right
] e^{-2\kappa_1(q,\omega)z} \label{Eq:s69}\\
\nonumber
G_{R,xx}(\mathbf{r},\mathbf{r},\omega)=G_{R,yy}(\mathbf{r},\mathbf{r},\omega)
&=&\frac{1}{8\pi}\int_0^{\infty}dq \frac{q}{\kappa_1(q,\omega)}
\left [
\frac{\kappa_2(q,\omega)-\kappa_1(q,\omega)}{\kappa_2(q,\omega)+\kappa_1(q,\omega)}\right
]
e^{-2\kappa_1(q,\omega)z}\\
&-&\frac{c^2}{8\pi\omega^2}\int_0^{\infty}dq \, \, q\kappa_1(q,\omega)
\left [
\frac{\epsilon_1\kappa_2(q,\omega)-\epsilon_2(\omega)\kappa_1(q,\omega)}{\epsilon_1\kappa_2(q,\omega)+\epsilon_2(\omega)\kappa_1(q,\omega)}\right
] e^{-2\kappa_1(q,\omega)z} \label{Eq:s68}
\end{eqnarray}\end{widetext}
where $\kappa_i(q,\omega)=\sqrt{q^2-\epsilon_i(\omega)
(\omega/c)^2}$ for $i=1, 2$.

All the parameters of the absorbing medium relevant to the SSQ
dissipative dynamics appear in the Green's tensor
$\widehat{\mathbf{G}}(\mathbf{r}_Q,\mathbf{r}_Q,\omega)$, where
$\mathbf{r}_Q$ denotes the SSQ location. The action of the
absorbing medium on a SSQ with dipole moment $\mathbf{p}$, is
completely described by the spectral density
\begin{eqnarray}
J(\omega)=\frac{1}{\pi\epsilon_0}\mathbf{p}\cdot\left [
\frac{\omega^2}{c^2}
Im[\widehat{\mathbf{G}}(\mathbf{r}_Q,\mathbf{r}_Q,\omega)]\right ]\cdot
\mathbf{p} \label{Eq:r21}
\end{eqnarray}
which is related with the qubit-environment coupling $g(\omega)$
and the density of states of the environment $\rho(\omega)$ by means
of $J(\omega)=g^2(\omega)\rho(\omega)$.
In order to compute the spectral function of a representative system,
we use parameters for silver in the range of frequencies of interest
where $\omega_p=3.76 eV$, $\epsilon_\infty =9.6$ and $\gamma_p=0.03
\omega_p$\cite{johnson72}, while for the dielectric constant at the
upper-half-space we take $\epsilon_1=5$. Since the density of states
has a singularity at $\omega_{sp}= \omega_{p}
\sqrt{\epsilon_{\infty} /(\epsilon_1 + \epsilon_{\infty})} =0.81
\omega_{p}$ non-Markovian effects associated with the structured
reservoir can be expected to occur around that inverse frequency.
Thus, we consider a SSQ with a dipole oriented along the
$z$-direction and an energy splitting $\omega_0$ slightly detuned
with respect to the singularity at the SP edge $\omega_{sp}$.
In particular, we calculate $J(\omega)$ for $\Delta=\bar{\omega}_{sp}
-\bar{\omega}_0=0.1$ where the frequencies have been normalized
to the bulk plasmon frequency $\bar{\omega}_{sp}=\omega_{sp}/\omega_p$
and $\bar{\omega}_0=\omega_0/\omega_p$.

In figure \ref{fig:density} we plot, for different values of the
qubit-surface renormalized separation $\bar{z}=z \omega_p/c$, the
spectral density in units of the spontaneous decay rate of the SSQ
in free space $\gamma_0=\omega_0^3 p^2/3 \pi \epsilon_0 c^3$ . Two
main results can be identified:

{\it i)} A strong reduction of $J(\omega)$ when the qubit gets
farther from the surface. This is a consequence of the exponential
reduction of the coupling, as a function of $z$, as indicated in
Eq.(\ref{Eq:s69}).

{\it ii)} For small separations $J(\omega)$ is highly structured
presenting a strong peak close to the frequency $\omega_{sp}$
(vertical line in Fig.(\ref{fig:density})). This is a consequence of
the singularity of the density of SP states at small detunings
$\Delta$. For increasing separation, $J(\omega)$ becomes much
smoother and a reduced maximum separates from $\omega_{sp}$.

\section{Time convolutionless method and non-Markovian effects\label{sec:tcl}}
The spectral density for the SSQ-SP system computed within a purely
classical scheme, can be used within a quantum framework describing
the dissipative dynamics of an open quantum system. As it is well
known, when the time correlation between the system and the
environment decay much faster than the characteristic inverse
dissipation rate, memory effects can be neglected in the so called
Markovian approximation, and the observables of the system are given
by analytical expressions. However, this timescale does not
represent the only one relevant to determine the system's evolution.
When the environment correlation time is longer than the inverse
rate of the system-bath coupling, new physics can arise at very
short times. A SSQ in close proximity to a dissipative metal
surface supporting SP modes sees a highly structured reservoir to
which it might be strongly coupled. Thus, one can expect
non-Markovian effects to be significant in the qubit time evolution.
Many
works\cite{chen09a,breuerbook02,lewenstein88,imamoglu94,stenius96,wilson02,breuer08,ma09,burkard09,chruscinski10,kaer10,roy10,haikka10,vacchini10}.
have been devoted to treat this problem at different levels of
precision and sophistication. Here, we chose to work within a
time-convolutionless (TCL) framework\cite {breuerbook02} to capture
non-Markovian effects to the lowest order in the SSQ-SP coupling
strength. This method has already been applied to consider the
spontaneous decay of a two-level system coupled to a general
structured reservoir\cite{breuerbook02}. For SSQ-SP systems, the
strong variation of $J(\omega)$ occurring for frequencies close to
$\omega_{sp}$, implies that dynamical features in time scales from
femtoseconds to picoseconds are expected.

\subsection{Time-convolutionless method}

What is of interest for us of the TCL method can be sketched as
follows\cite{breuerbook02}. It consists in transforming the typical
non-Markovian integro-differential equation for the reduced density
matrix into a local in time evolution equation by making use of a
power expansion technique of the Nakajima-Zwanzig type. As a result,
a master equation for a qubit is obtained with time dependent decay
rate $\gamma (t)$ and Lamb shift $S(t)$:
\begin{eqnarray}
\label{mastertcl}
& &\frac{d\rho(t)}{dt}=\frac{i}{2}S(t)[\rho(t),\sigma^+\sigma^-]\nonumber\\
& & + \frac{\gamma(t)}{2}(2\sigma^{-}
\rho(t)\sigma^{+}-\sigma^{+}\sigma^{-}\rho(t)-\rho(t)\sigma^{+}\sigma^{-})
\label{Eq:me1}
\end{eqnarray}
Time-dependent rates can be calculated within a perturbative
expansion. In order to calculate them, a first step is to  Fourier
transform the spectral density:
\begin{eqnarray}
 \label{fouriert}
f(t)=\int d\omega J(\omega) e^{i(\omega_0 -\omega)t}
\end{eqnarray}
The lowest order non-Markovian effects, i.e. the so-called
post-Markovian behavior, are contained in the second order
contributions to $\gamma(t)$ and $S(t)$ given by:
\begin{eqnarray}
\gamma_2(t)=\frac{1}{2}\int_0^t dt_1 \Re f(t-t_1) \label{Eq:col1}\\
S_2(t)=\frac{1}{2}\int_0^t dt_1 \Im f(t-t_1).
\end{eqnarray}
where $\Re$ and $\Im$ denote real and imaginary parts,
respectively.

\subsection{SSQ spontaneous decay}

We start by considering the situation where an undriven SSQ is
prepared in its excited state from which it decays emitting a photon
to the vacuum or to the SP field. There are two possible situations
depending on the sign of the detuning $\Delta$ between SP
$\bar{\omega}_{sp}$ and the SSQ $\bar{\omega}_0$ renormalized
frequencies. The time evolution of the excited state population, is
given by\cite{breuerbook02}
\begin{equation}
 \label{populationnolaser}
n_1(t)=n_1(0) e^{-\int_0^t \gamma_2(s) ds},
\end{equation}
with $n_1(0)=1$ and the decay rate obtained from
Eq.(\ref{Eq:col1}).

At this stage, we want to analyze the importance of memory effects.
Therefore, in the calculations reported in this subsection we
do not include the free space part of the Green's tensor,
Eq.(\ref{Eq:s17}), which involves a much slower dynamics than the
one associated to the reflection contribution
$\widehat{\mathbf{G}}_R$, as given in Eq.(\ref{Eq:s69}).

\begin{figure}[ht]
  \centering
\includegraphics[width=0.99\linewidth]{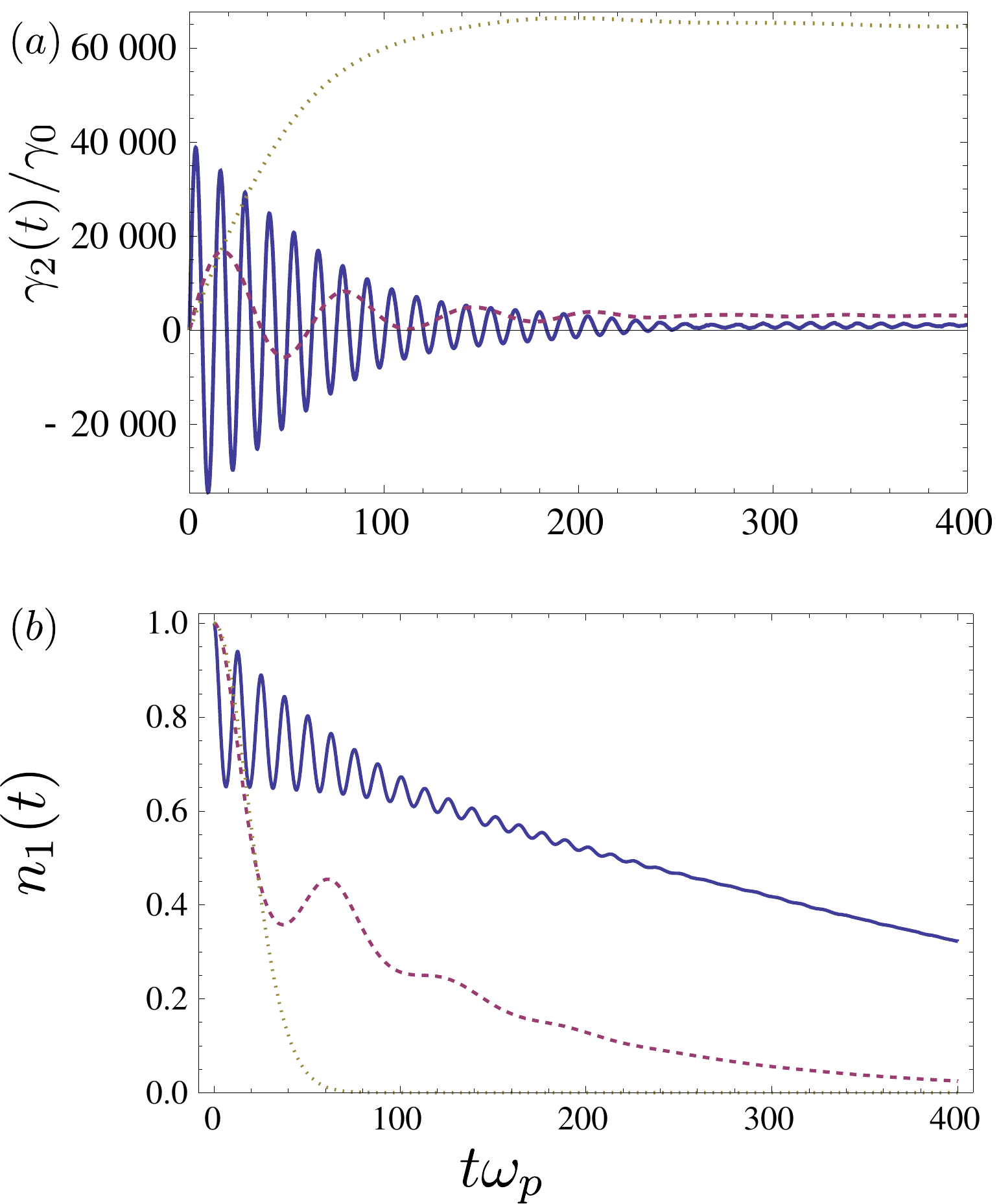}
\caption{Decay rate (a) and population of the excited state (b) of a SSQ
located at $\bar{z}=0.055$ from the planar surface. Different lines
correspond to different detunings from the SP frequency:
$\Delta=\bar{\omega}_{sp}-\bar{\omega}_0=0.5$(solid blue),
$0.1$(dashed purple) and $0.01$(dotted yellow).}
  \label{fig:popinside}
\end{figure}

Figure \ref{fig:popinside} shows $n_1(t)$ for different positive
detunings, i.e. when the SSQ is resonant with the continuum stripe
of SP modes($0<\omega<\omega_{sp}$). In order to have a highly
structured reservoir, we have taken a small qubit-surface
separation, $\bar{z}=0.055$ (very close to the blue line spectral
density in figure \ref{fig:density}). For large detuning
$\Delta=0.5$, $\gamma_2(t)$ oscillates around a constant (Markovian)
value. At some time intervals, $\gamma _2 (t)$ takes on negative
values, a fact that tends to slow down the decay of the excited
state population. Physically, this behavior can be understood as due
to the back-action of the reservoir on the SSQ re-exciting it.
When the SSQ splitting energy gets closer in resonance with the top
SP energy, e.g. $\Delta=0.1$, the oscillations slow down, the envelope
of the oscillatory decay rate becomes smaller and the negative
parts of the decay rate $\gamma_2(t)$ tend to vanish producing only
few oscillations before the spontaneous decay becomes almost
exponential. For further smaller detunings, e.g. $\Delta=0.01$, the SSQ sees
an even more structured reservoir with a decay rate modifying
completely its behavior: its value increases considerably and it
just oscillates slightly around a large positive value, producing a
monotonous decay of the SSQ excited state population. This last result
indicates that the second-order TCL method is approaching its limit of
validity. Physically, this behavior is a consequence of the fast
transfer of the SSQ energy to the SP field, an energy which is
irreversibly lost and the quantum emitter ends up in its ground
state.

\begin{figure}[ht]
\includegraphics[width=0.99\linewidth]{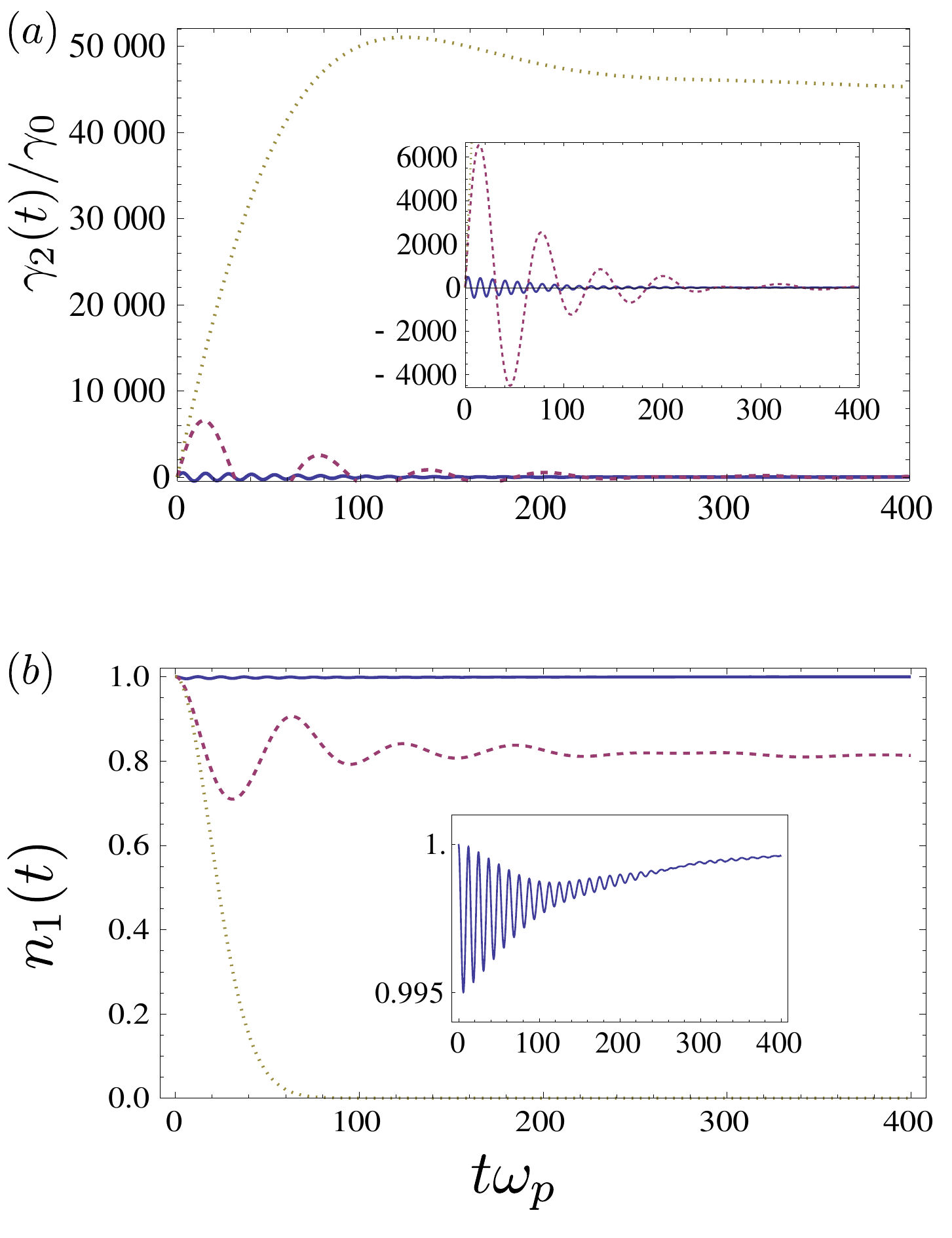}
   \caption{Decay rate and population of the excited state for a SSQ placed at $\bar{z}=0.055$
from the planar surface. The different lines correspond to different
detunings: $\Delta=\bar{\omega}_{sp}-\bar{\omega}_0=-0.5$(solid
blue), -$0.1$(dashed purple) and -$0.01$(dotted yellow). Inset in (a) corresponds to a zoom of the decay rate for the first two values of $\Delta$, while inset in (b) corresponds to a zoom to population of the excited state at short times for $\Delta=-0.5$. Including the effect of free space emission ($\gamma_0$) produces a decay
of $n_1$ in a time scale much larger than that of the figure.
}
  \label{fig:popoutside}
\end{figure}

On the other hand, a physically different situation occurs for
negative detunings, i.e. when the SSQ energy falls within the gap
where no SP states exist. Figure \ref{fig:popoutside} is similar to
figure \ref{fig:popinside} with the same $\bar{z}$ and detunings
with just a change of sign with respect to the ones in figure
\ref{fig:popinside}. When ${\omega}_0$ is far above the edge of
the SP dispersion, the SSQ basically remains in its excited state
as the spectral density for these energies is practically zero,
so there are not accessible SP modes to which decay into.
Nevertheless, as shown in the inset of
figure \ref{fig:popinside}, one may observe some non-Markovian
oscillations for very short times. When the SSQ energy is tuned
closer in resonance with $\bar{\omega}_{sp}$, e.g. $\Delta=0.1$, a
very interesting phenomenon occurs: the emitter undergoes the
so-called fractional decay in which the population tends to a
finite, non-zero, value at long times. Including the effect of free
space emission ($\widehat{\mathbf{G}}_0$) produces a decay of $n_1$ in a time
scale ($\gamma_0^{-1}$) much larger than that of the figure. As the emission
frequency is further scanned closer to the band edge, the behavior
changes again dramatically: the decay rate, instead of oscillating
around zero, oscillates slightly around a positive value, which
results into an irreversible exponential decay.

A very important result must be drawn from all these results: the
timescale of these non-Markovian effects is a few hundred times
$\omega_p ^{-1}$. For normal metals, this means times below $1ps$.
Beyond that short timescale, $\gamma_2$ becomes constant just at the
value it takes in a Markovian description as discussed in the
following Section IV.

\subsection{Coherently driven SSQ}

\begin{figure}[ht]
\includegraphics[width=0.99\linewidth]{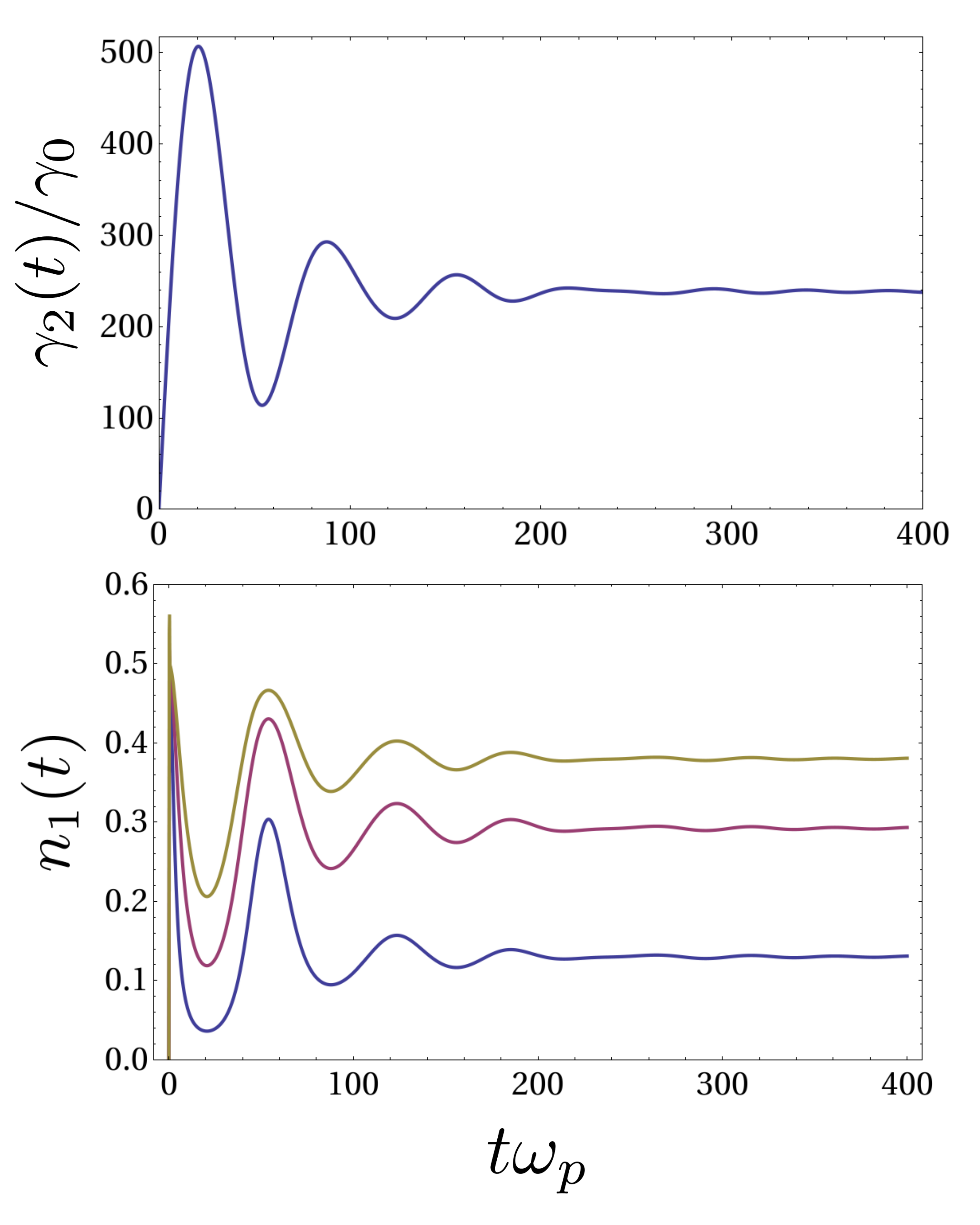}
   \caption{Decay rate and excited population of a SSQ placed
at $\bar{z}=0.2$ with energy detuning
$\Delta=\bar{\omega}_{sp}-\bar{\omega}_0=0.1$ and coherently driven
by a laser in resonance with the SSQ. The different lines correspond
to different laser intensities
$\Omega/\gamma_0=100$(blue), $200$(purple) and $500$(yellow)}
  \label{fig:poplaser}
\end{figure}

After having studied the effect of the structured reservoir on the
SSQ spontaneous emission, now we turn our attention to the case
where the system is coherently driven by a laser field. The SSQ
emits and absorbs photons simultaneously. The system can achieve a
stationary state in which light absorbed from the laser ends being
transferred to plasmons. The laser can be treated as a classical
field included in the, local in time, coherent part of the master
equation (\ref{mastertcl}) through the Hamiltonian
$H_{\mathrm{las}}=\Omega(\sigma^+ e^{i \omega_{\mathrm{las}}t}
+e^{-i \omega_{\mathrm{las}}t} \sigma^-)/2$. The Rabi frequency
$\Omega$ measures the strength of the coupling of the SSQ dipole
with the laser field. The time evolution of the SSQ
excited state population is obtained from the solution of
Eq.(\ref{Eq:me1}) represented in a rotating frame at the laser
frequency $\omega_{\mathrm{las}}$. $n_1 (t)$ is plotted in figure
\ref{fig:poplaser}, for the case of perfect
resonance between the laser and the SSQ. We assume that
$\gamma_2 (t)$ remains unaffected by the laser field, so that the
main effect of the laser is to bring the system to a
stationary state in a timescale which is similar to that of the
spontaneous decay discussed in the previous subsection.

\section{Optical properties in the Markov approximation\label{sec:Markov}}

The results of the previous section show that a SSQ presents significant
non-Markovian effects in a timescale a couple of orders of magnitude
larger than $\omega_p ^{-1}$. Hereafter, we concentrate in the usual case
of having a resolution in time larger than a picosecond.
Then, the system can be described by a Markovian dynamics given by a
Master equation like Eq. (\ref{mastertcl}) but now with a Lamb-shift
$S=S_2(t\rightarrow \infty)$ and a constant
decay rate $\gamma=\gamma_2(t\rightarrow \infty)$
including both the free space and the reflection contributions
to the dissipative dynamics. Since the only effect of the
Lamb-shift is a constant energy shift, from now on we do not pay
attention to it.

\subsection{Decay rate}
The decay rate in the Markovian regime coincides with the long
time limit of  $\gamma_2(t)$, $\gamma=\gamma_2(t\rightarrow
\infty)$, allowing to identify $\gamma$ as simply the spectral
function at the SSQ frequency:
\begin{eqnarray}
\label{decay}
\gamma=2\pi J(\omega_0)=\frac{2\omega_0^2}{\epsilon_0 c^2}\mathbf{p}\cdot
Im[\widehat{\mathbf{G}}(\mathbf{r}_Q,\mathbf{r}_Q,\omega_0)]\cdot
\mathbf{p}  \label{Eq:s12}
\end{eqnarray}
where the two terms corresponding to the free space
($\widehat{\mathbf{G}}_0$) and the reflection part
($\widehat{\mathbf{G}}_R$) of the dissipative dynamics are
included in $\widehat{\mathbf{G}}$.

\begin{figure}[ht]
\includegraphics[width=0.99\linewidth]{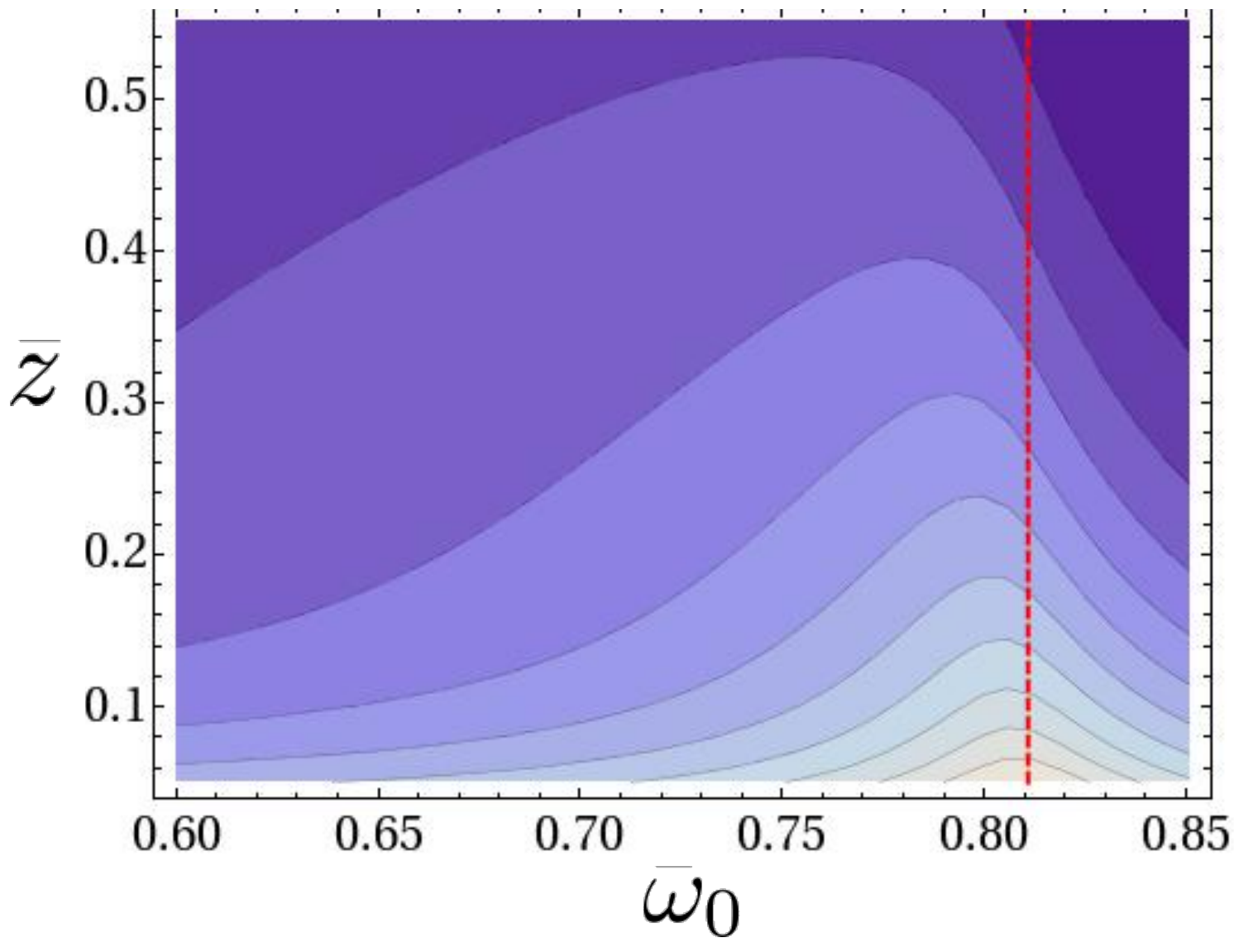}
\caption{Surface assisted decay rate $\log_{10}\left ( \gamma/\gamma_0\right )$ given
by Eq. (\ref{decay}) for a region of the
$\bar{z} , \bar{\omega}_0$ parameter space. The red dotted
gridline marks the $\bar{\omega}_{\mathrm{sp}}$ frequency.
Lighter blues correspond to high values. Between the highest and the
lowest values there are four orders of magnitude. }
  \label{fig:gammaMark}
\end{figure}

\begin{figure}[ht]
 \centering
\includegraphics[width=0.99\linewidth]{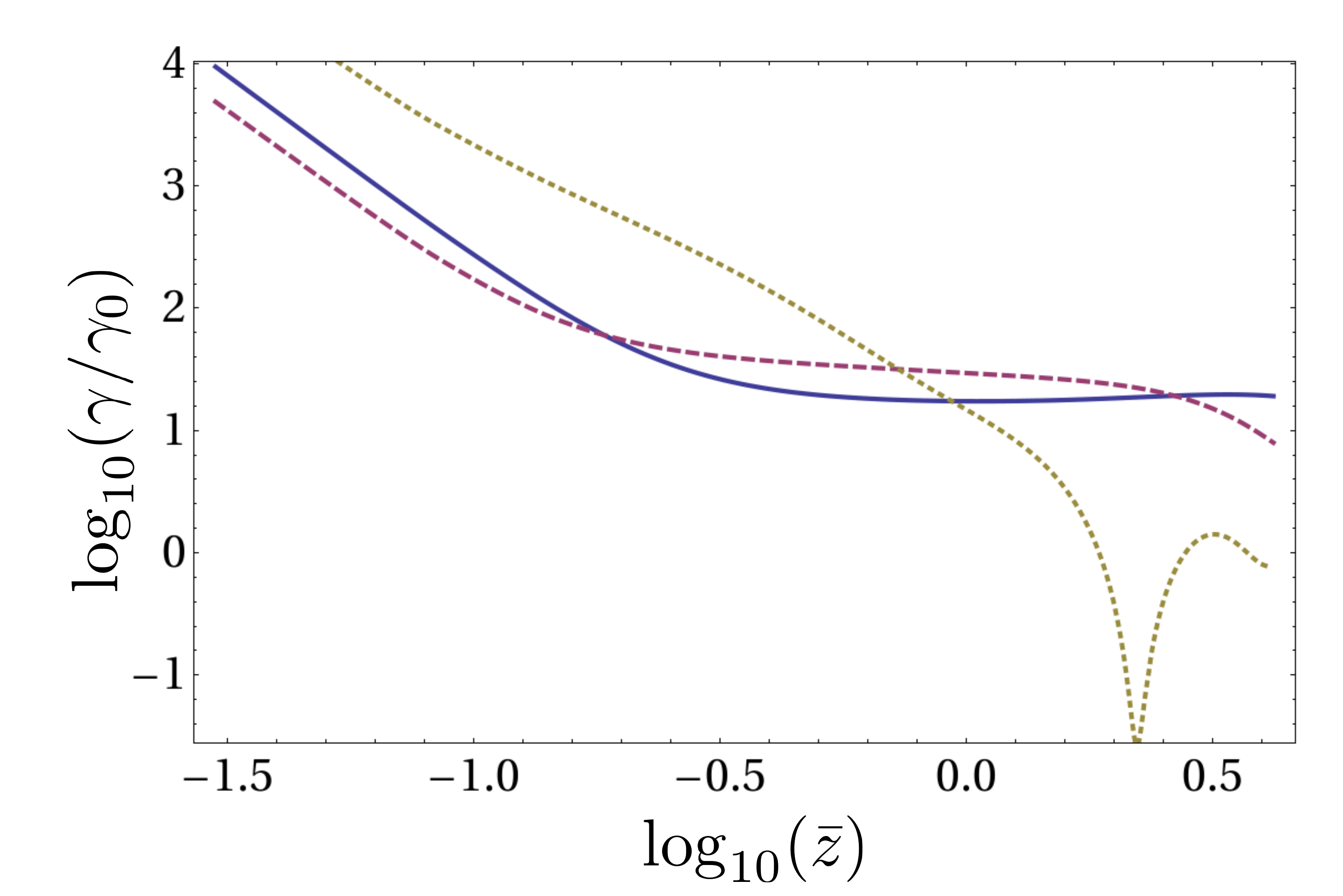}
\caption{Logarithmic relative decay rate $\log_{10}(\gamma /
\gamma_0)$ calculated with Eq.(\ref{decay}) as a function of
$\log_{10}(\bar{z})$. The three curves correspond to different SSQ
energies $\bar{\omega}_{0}= 0.24$(solid blue), $0.4$(dashed purple)
and $0.76$ (yellow dotted).}
    \label{fig:figdecay}
 \end{figure}

The SSQ decay rate to the SP reservoir of the metallic surface,
$\gamma$, is shown in figure \ref{fig:gammaMark}, in a parameter
space $\{\bar{z} , \bar{\omega}_0\}$, where lighter blues correspond to
high decay rates with a variation of four orders of magnitude
between the highest and the lowest values. In order to discuss these
results, it is better to plot $\gamma$ vs. the SSQ-interface
distance $\bar{z}$ (in logarithmic scales) for different SSQ
energies as depicted in figure \ref{fig:figdecay}. It is worth
noticing two important features: First, at a large $\omega_0$ value
the assisted decay rate is smaller than the vacuum one for a certain
range of distances, due to the fact that the reflected part of the
Green's tensor is interfering destructively with the direct one.
This effect is evident when the SSQ frequency approaches
$\omega_{sp}$ while it moves to larger separations $\overline{z}$,
and it weakens, when $\omega_0$ is far from the SP band edge.

\begin{figure}[ht]
 \centering
\includegraphics[width=0.99\linewidth]{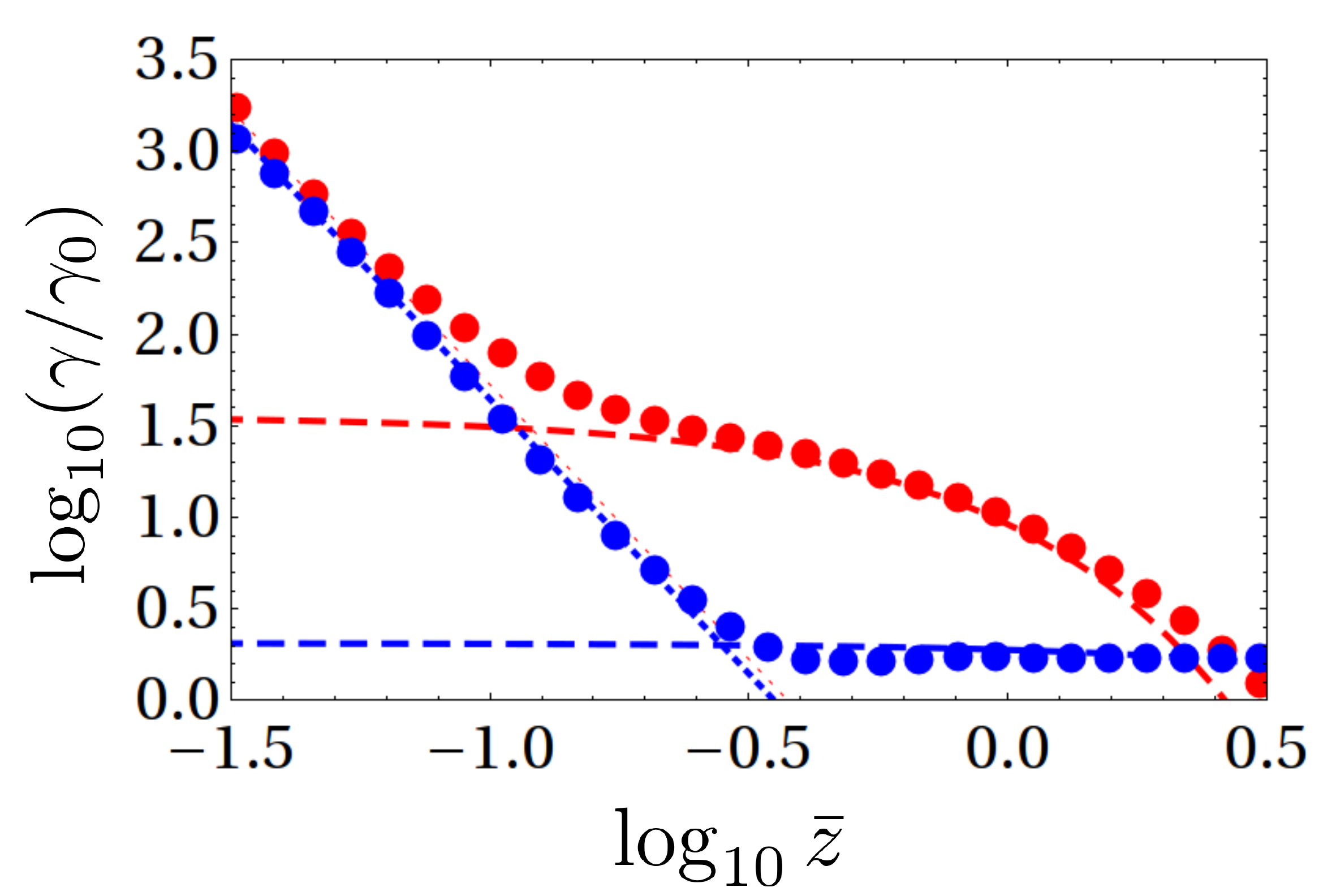}
\caption{Logarithmic relative decay rate $\log_{10}(\gamma
/\gamma_0)$ calculated with Eq.(\ref{decay}) as a function of
$\log_{10}(\bar{z})$ with $\epsilon_1=\epsilon_{\infty}=1$ and
$\gamma_p=\omega_p/500$. Two different SSQ energies are
considered: $\bar{\omega}_0=0.2$ (blue circles, far from the SP
edge) and $0.6$ (red circles, close to the SP edge). Dashed lines
denote the decay rates as predicted by a single pole approximation
(SP contribution). Dotted lines correspond to metal losses as the
emission of electron-hole pairs.}
    \label{fig:figdecaysep}
\end{figure}

Second feature is even more important. When the SSQ-surface distance
varies, the decay rate suffers a transition from a $1/z^3$ behavior
to a much slower decrease. In order to understand the physics behind
this behavior, we analyze the much simpler case
$\epsilon_1=\epsilon_{\infty}=1$ and $\gamma_p=\omega_p/500$, i.e.
a rate for losses one order of magnitude smaller than the one we
have used in previous cases. Large dots in figure \ref{fig:figdecaysep}
depict the decay rate as calculated with the full Green's tensor for
two different SSQ frequencies.
At very short distances the decay rate is highly enhanced with respect
to $\gamma_0$ and it shows a $1/z^3$ dependence, which can be
obtained (dotted lines in the figure) by means of a model which only contains
non-radiative processes as the creation of electron-hole pairs
in the metallic medium. A crossover to a different behavior occurs at
a critical distance, which depends on the SSQ-SP detuning $\Delta$.
Beyond this critical distance, a single plasmon pole approximation
(dashed lines in the figure) in the Green tensor in Eq.(\ref{Eq:s12})
is able to reproduce the numerical result with the complete Green's
tensor. In other words, for separations beyond the crossover,
the SSQ decay just produces the emission of surface plasmons while
other metallic losses become negligible. In the case of the SSQ embedded
in a dielectric or a metal with very large
losses this crossover can be hindered by other physical effects
such as those coming from local dissipative circulating
currents\cite{dzsotjan10a}. As the SP channel contribution increases
when the SSQ energy gets closer to the plasmon band edge, this
crossover effect can be exploited in designing coherent plasmonic
devices\cite{vasa08a}.

\subsection{Spectrum and correlation functions}

\begin{figure}[!ht]
\includegraphics[width=0.99\linewidth,height=18cm]{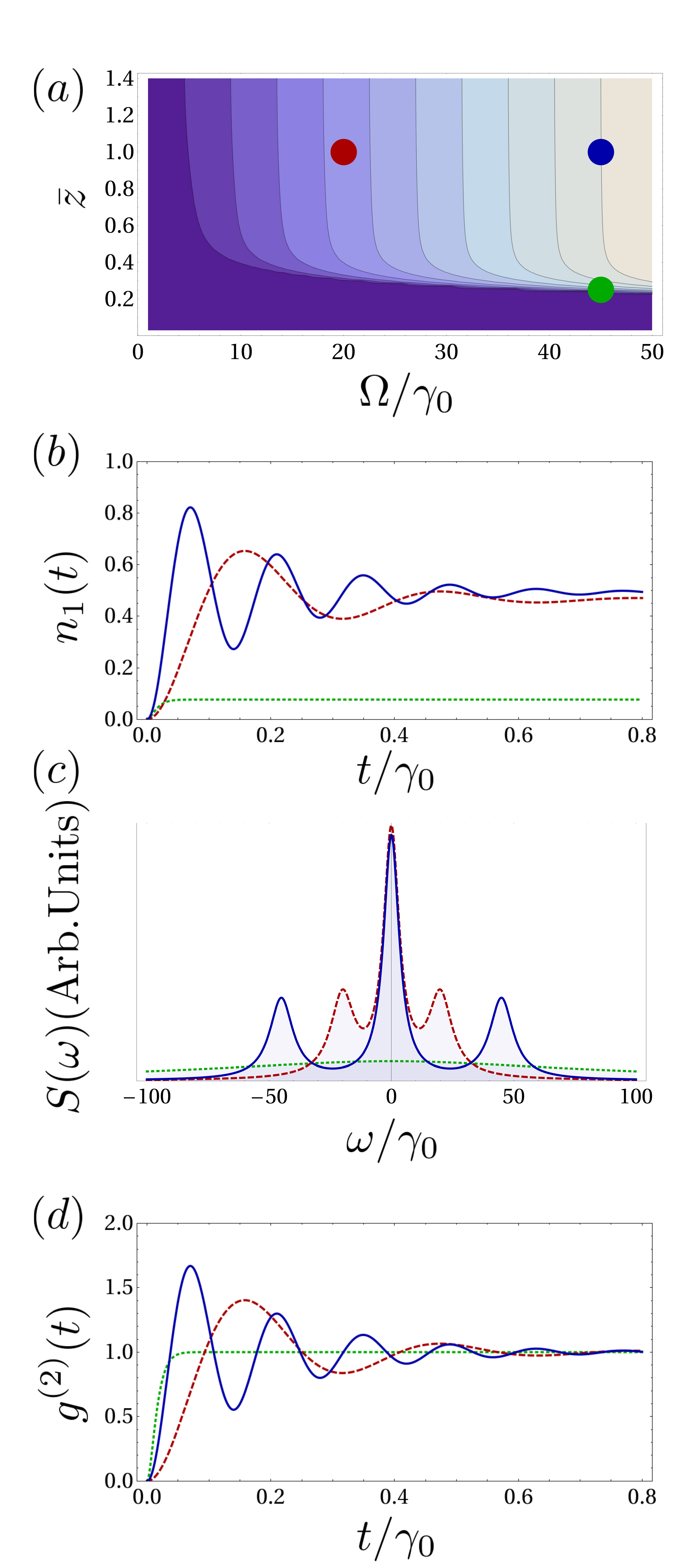}
\caption{Optical properties of the SSQ-SP system. Panel (a)
corresponds to the value of $\Re (R)$ in the parameter space
$\{\bar{z} , \bar{\Omega}_0 \}$ in order to distinguish the strong
and weak coupling regions. Panel (b) and (d) show the dynamics of
the excited state population and the two-photon correlation function
($g^{(2)}(t)$) respectively for the three points plotted in panel (a).
Panel (c) shows the qubit luminescence spectra for those three
particular cases.}
  \label{fig:markov}
\end{figure}

\begin{figure}[!ht]
\includegraphics[width=0.99\linewidth,height=18cm]{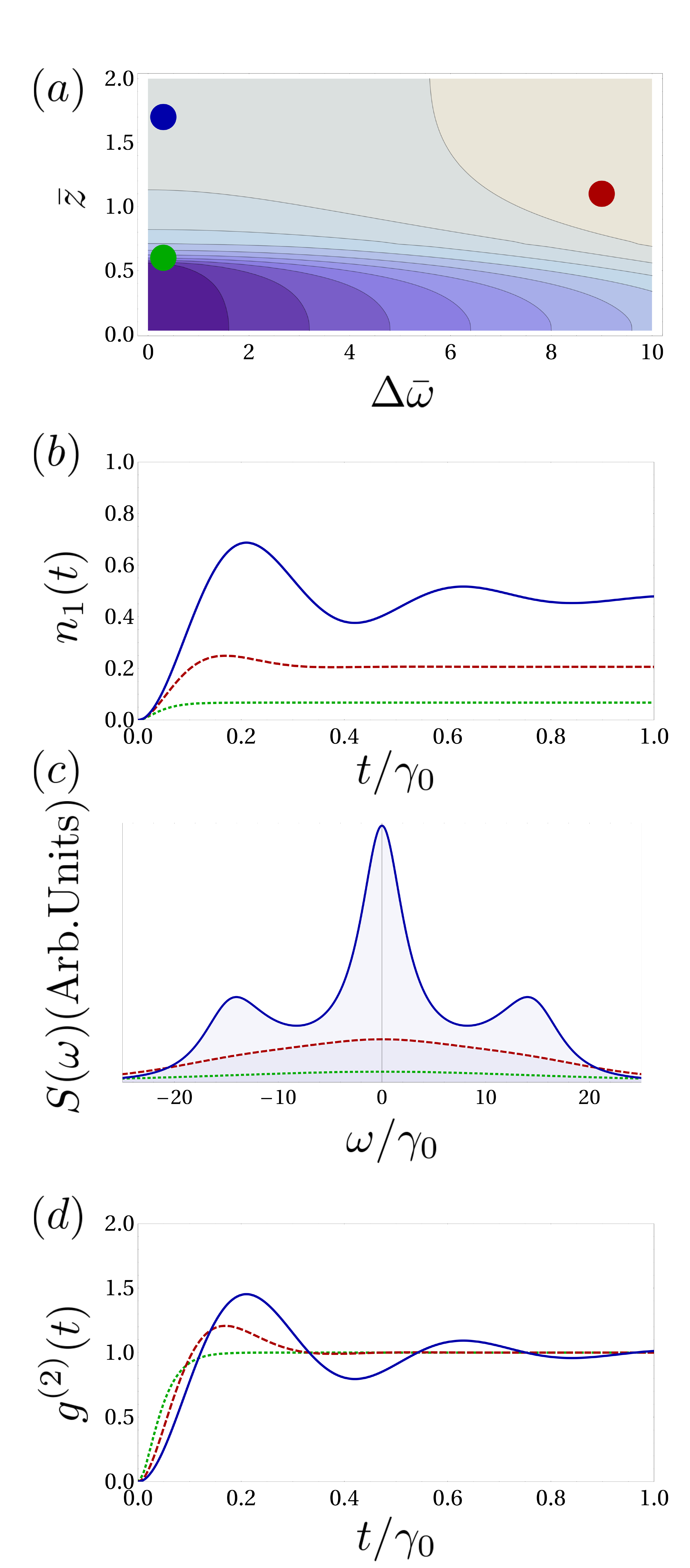}
\caption{Optical properties of the SSQ-SP system. Panel (a)
corresponds to the value of $\Re (R_\Delta)$ in the parameter space
$\{\bar{z} , \Delta\bar{\omega}\}$ to distinguish the strong and
weak coupling regions. Panel (b) and (d) show the dynamics of the
excited state population and the two-photon correlation function
($g^{(2)}(t)$) respectively for the three points plotted in panel (a).
Panel (c) shows the SSQ luminescence spectra for those three
particular cases.}
  \label{fig:markovOUT}
\end{figure}

Hereafter, we consider the case in which the system is coherently
driven by a laser so that the system reaches a stationary
state with partial occupation of the two levels of the SSQ.
The master equation in the rotating frame at the
laser frequency, $\omega_{\mathrm{las}}$, is:
\begin{eqnarray}
\label{master}
\frac{d\rho(t)}{dt}=i \{ \Delta \omega  [\rho(t),\sigma^+\sigma^-]+ \frac{\Omega}{2} [\rho(t),\sigma^+ +\sigma^-]\} \nonumber\\
+ \frac{\gamma}{2}(2\sigma^{-}
\rho(t)\sigma^{+}-\sigma^{+}\sigma^{-}\rho(t)-\rho(t)\sigma^{+}\sigma^{-})
\end{eqnarray}
with $\Delta \omega=\omega_0-\omega_{\mathrm{las}}$. The problem
reduces to the study of the SSQ resonance fluorescence
\cite{walls94a} near the planar surface of a dissipative metal. The
vacuum resonance fluorescence has been widely studied in the
literature in the case of resonant excitation for which an
analytical solution for the population, spectrum and second order
coherence function exists. Here, we extend such analysis to the
non-resonant case and pay special attention to the effect of the SP
reservoir. We present our analysis in the way the experiments can be
performed either by tuning the laser resonantly with the SSQ energy
and varying the laser intensity or by fixing the laser intensity and
scanning the laser frequency. Figure \ref{fig:markov} explores the
former alternative whereas in figure \ref{fig:markovOUT} we consider
the later one.

From the master equation (\ref{master}) one may derive the equations
of motion for the expectation values $\langle \sigma^+(t) \rangle$,
$\langle \sigma^-(t) \rangle$ and $\langle \sigma^z(t) \rangle$
arriving to the well-known optical Bloch equations(OBE).  The
steady-state solution for the excited state population is:
\begin{eqnarray}
 \label{steadypop}
\langle n_1 \rangle_{ss}=\frac{\Omega^2}{(\gamma^2+4\Delta\omega^2+2 \omega^2)}.
\end{eqnarray}
In general, the OBE must be solved numerically in order to get the
population dynamics $\langle n_1(t) \rangle$, except for the
resonant case ($\Delta\omega=0$) for which an analytical solution
exists:
\begin{eqnarray}
\label{timepop}
 & & \langle n_1(t) \rangle=\frac{\Omega^2}{\gamma+2\Omega^2} \nonumber \\ & & \times \left[ 1-e^{-3\gamma t/4} \left(\cos(R t ) +\frac{3\gamma}{4 R}\sin(R t)\right)\right].
\end{eqnarray}
where $R=\sqrt{\Omega^2- \gamma^2/16}$, labeled as Rabi
splitting at resonance\cite{khitrova}, is the parameter
characterizing the strength of the effective coupling. There is a
threshold for the laser intensity at $\Omega=\gamma/4$. For $\Omega$
below this threshold, the solutions are monotonically decaying functions of time
so that the system is said to be in the weak coupling (WC) regime.
Above that threshold, the populations exhibit oscillations, and the
system is said to be in the strong coupling (SC) regime. In figure
\ref{fig:markov}(a) we plot the real part of $R$ in the parameter
space $\{\bar{z} , \bar{\Omega}_0 \}$. The bluest region corresponds
to $\Re (R)=0$, which means that the Rabi splitting at resonance is
purely imaginary and consequently the system is in the WC regime.
For the regions in which blue becomes lighter, the values correspond
to positive and higher values of $\Re (R)$.

In order to clarify these results, we show the population dynamics
in figure \ref{fig:markov}(b) for three different points highlighted
in part (a) of the same figure: the green curve corresponds to a
configuration where the laser is weakly coupled to the system, so no
oscillations are observed in the population. The red point
corresponds to the region of transition from WC to SC where just one
clear oscillation occurs before practically arriving to the steady
state. Finally the blue point corresponds to a configuration where
the laser is strongly coupled to the SSQ and several oscillations
are observed before the steady state is achieved.

Another experimental alternative is to keep $\Omega$ constant and
vary the laser frequency as it is plotted in figure
\ref{fig:markovOUT}. In this case, the laser is out of resonance and
the Rabi splitting must be redefined as\cite{khitrova}:
\begin{eqnarray}
\label{rabidelta}
R_\Delta=\sqrt{\Omega^2-\left(\frac{\gamma}{4}+i \Delta \omega\right)^2}
\end{eqnarray}
Figure \ref{fig:markovOUT}(a) shows $\Re R_{\Delta}$ in the
parameter space $\{\bar{z} , \Delta\bar{\omega} \}$. The population
dynamics is shown in panel (b) of the same figure, where one can
observe the transition from strong coupling (oscillations, solid
blue) to weak coupling (monotonous in time, green dotted) for
non-resonant excitation of the SSQ.

A clear manifestation of the transition from WC to SC appears in the
optical spectrum at the stationary regime. It can be calculated
through the Wiener-Khintchine formula:
\begin{eqnarray}
 \label{spectrum}
S(\omega))=\frac{1}{ \pi} \Re \int_{0}^{\infty}e^{i \omega \tau}\langle\sigma^+(t)\sigma^-(t+\tau)\rangle d\tau.
\end{eqnarray}
The calculation of the two-time correlator in Eq. (\ref{spectrum})
requires the use of the Quantum Regression Theorem\cite{walls94a} by
using the steady state populations as initial values for the second
time dynamics. In the resonance fluorescence problem there are
always two main contributions to the spectra: the Rayleigh
scattering coherent part and the one coming from the incoherent
scattering. The former contribution is just a delta function at
$\omega_0$ that we ignore in our results. We are mainly interested
in the contribution coming from the inelastic scattering which is
shown in figures \ref{fig:markov}(c) and \ref{fig:markovOUT}(c) for
the resonant and non-resonant cases, respectively. As it occurred
with the population, under resonant excitation the spectrum admits
an analytical expression:
\begin{eqnarray}
 \label{spectrarev}
& S(\omega)\propto & \Big(\frac{\gamma/4}{\frac{\gamma^2}{4}+(\omega-\omega_0)^2}
 +\frac{3\gamma/16}{\frac{9\gamma^2}{16}+(\omega-\omega_0+\Omega)^2} \nonumber\\
& & + \frac{3\gamma/16}{\frac{9\gamma^2}{16}+(\omega-\omega_0-\Omega)^2} \Big)
\end{eqnarray}
In the WC regime (green point) $\Omega<\gamma/4$, the light emitted
simply produces a Lorentzian curve peaked about $\omega_0$ with
linewidth $\gamma/2$. For the intermediate regime (red point), on
top of the Lorentzian peaked at the qubit frequency, some satellites
start to appear at the laser Rabi frequency $\pm \Omega$. For a
strong-driving field situation $\Omega>\gamma/4$ these two sidebands
appear at frequencies $\omega=\omega_0\pm\Omega$. For the
non-resonant case, the threshold changes but the behavior remains
qualitatively unaffected: even though the dressed state structure is
slightly modified by the detuning, at the end, a triplet is obtained
in the resonant case. The existence of this Mollow's triplet is a
manifestation of the SC of the laser to the SSQ-SP system.

Another magnitude of experimental interest is the second order coherence function:
\begin{eqnarray}
\label{secondorder}
g^{(2)}(\tau)= \frac{G^{(2)}(t,t+\tau)}{G^{(1)}(0)G^{(1)}(\tau)}
\end{eqnarray}
with correlation functions
\begin{eqnarray}
& G^{(2)}(t,t+\tau)= & \langle \sigma^{(+)}(t) \sigma^{(+)}(t+\tau) \sigma^{(-)}(t+\tau) \sigma^{(-)}(t)\rangle
\nonumber \\
& G^{(1)}(t,t+\tau)= & \langle \sigma^{(+)}(t) \sigma^{(-)}(t+\tau) \rangle .
\end{eqnarray}
We evaluate these magnitudes at the stationary state. In the resonant case, the
second order coherence function can be analytically expressed as:
\begin{eqnarray}
\label{secondordertwolevelsystem} g^{(2)}(\tau)= 1-e^{-3 \gamma
\tau/4}\left(\cos(R \tau) + \frac{3\gamma}{4 R}\sin(R\tau)\right).
\end{eqnarray}
It clearly exhibits photon anti-bunching: $g^2(0)=0$. Figure
\ref{fig:markov}(d) shows $g^{(2)}$ for zero detuning for the three
different points considered above for the other magnitudes. Apart
from the antibunching, the case of SC shows a remarkable oscillatory
behavior. Once more, qualitatively similar results are obtained with
laser-SSQ detuning as shown in figure \ref{fig:markovOUT}(d).

The main consequence to be drawn from figures \ref{fig:markov} and
\ref{fig:markovOUT} is that by pumping the SSQ with a tunable laser,
and measuring spectra and second order correlation functions, one
can extract information about the SSQ coupling to the surface
plasmon of the dissipative metal.

\section{Summary\label{sec:Summary}}

In this work we have studied the properties of the coupling of light
with a SSQ, embedded in a dielectric, in the presence of a SP field
supported in the interface between this dielectric matrix and a
dissipative metal. Using a time-convolutionless approach, we provide
a theoretical description of the non-Markovian features for this
kind of systems and discuss its relevance in possible
observations. In a spontaneous decay situation, different behaviors
occur depending on both the sign and the absolute value of the
SSQ-SP detuning: from a monotonous (almost exponential) decay for
very small detunings, to population oscillations due to
reabsorptions in the case of positive detuning. Even fractional
decays can be observed, when negative detunings are present and the
SSQ energy is not too close to the SP edge band.

In experimental situations, non-Markovian features can be hard to
detect due to practical difficulties in getting the adequate
time-resolution. Therefore, we have also considered a Markov
approximation to study the electrodynamics of the SSQ coupled to a
reservoir of SP modes. The whole information of the planar metallic
surface is embedded in the decay rate constant, which depends on
both the SSQ frequency and distance to the surface. The excitation
of the system by a laser allows the existence of a steady state as
well as the analysis of different measurable properties of the
SSQ-SP system as, for instance, surface enhancements of rate
emission, optical spectra and time-dependent photon-photon
correlation functions. Our main result is that the qubit decay shows
a crossover passing from being purely dissipative for
small qubit-surface distances to plasmon emission for larger
separations. As the SP emission channel increases
when the SSQ energy gets closer to the plasmon band edge, this
crossover effect can be exploited in designing coherent plasmonic
devices. Our next task, beyond the scope of the
present work, is to treat the plasmonic part of the system not as a
reservoir but as an ingredient coherently coupled to one or more
SSQs\cite{vasa08a}.

\section{Acknowledgements\label{sec:acknowledgements}}

Helpful discussions with F.J. Garcia-Vidal and L. Martin-Moreno are
acknowledged. Work supported in part by the Spanish MEC under
contracts Consolider-Ingenio2010 QOIT-CSD2006-00019 and
MAT2008-01555, and by the CAM under contract S-0505/ESP-0200. A.G-T
acknowledges funding from  AP2008-00101 grant of Spanish Education
Ministry. L.Q. was partially supported by Faculty of
Sciences-Research Funds 2009 (UniAndes). F.J.R. was partially
supported by Banco de la Republica (Colombia).

\bibliography{plasmon,nonmarkovian}



\end{document}